# Temperature-driven phase transitions in $SrBi_2Ta_2O_9$ from first-principles calculations

*R. Machado, M. Sepliarsky and M.G. Stachiotti*

*Instituto de Física Rosario, Universidad Nacional de Rosario,*
*27 de Febrero 210 Bis, (2000) Rosario, Argentina.*

**Abstract**

The phase transition sequence of $SrBi_2Ta_2O_9$ is investigated using a shell model with parameters fitted to first-principles calculations. We show that the complex interplay between polar and nonpolar instabilities leads to the presence of two phase transitions, corroborating the existence of an intermediate orthorhombic paraelectric phase. This phase is characterized by the rotation of the $TaO_6$ octahedra around the a-axis. We show that this phase can be also detected from the dielectric response of the material. The present approach constitutes a powerful tool for a theoretical prediction of intermediate phases, not yet observed experimentally, in other Aurivillius compounds.

In the last decade, many ferroelectric materials have been studied for application in ferroelectric random access memory (FeRAM) devices. $SrBi_2Ta_2O_9$ (SBT) and $Bi_4Ti_3O_{12}$ (BIT) are very attractive for FeRAM development as they are distinguished by their capabilities for low voltage operation and good reliability properties [1]. These materials, which belong to the family of Aurivillius compounds with a general formula $Bi_2O_2(A_{m-1}B_mO_{3m+1})$, are layered bismuth oxides formed by stacking $Bi_2O_2$ slabs and perovskitelike blocks. Their high-temperature paraelectric phase has *I*4/*mmm* tetragonal symmetry. They are ferroelectric at room temperature, but in addition to the polar distortion, they have other nonpolar antiferrodistortive distortions, which involve large tiltings of the perovskitelike octahedra. Understanding the properties of Aurivillius compounds is important for future device applications but presents a challenge to theory and modelling because of the complex interplay between the different structural instabilities, which suggests that the phase transition sequence in these materials is not as simple as considered until now [2].

In SBT, for example, three main distortions from the tetragonal structure lead to the ferroelectric *A2₁am* phase: the ions displace along the orthorhombic *a*-axis (the [110] direction of the tetragonal structure), and $TaO_6$ octahedra rotate around the *a* and *c* axes [3]. The first factor is directly responsible for the macroscopic spontaneous polarization along the *a* direction, which mainly involves displacements of the $Bi_2O_2$ planes relative to the perovskitelike blocks [4]. Recent ab-initio calculations demonstrated that a delicate interplay exists between these three structural instabilities, and the ferroelectric phase transition was explained by coupling the ferroelectric soft mode with another hard vibration associated with the octahedral tilting at the Brillouin zone boundary [5]. In this respect, the nature and sequence of phase transitions in SBT is much more complicated as compared to simple perovskites.

For a long time, an intermediate phase between the high-symmetry tetragonal and the ferroelectric phase has been suggested for some Aurivillius compounds [6]. These results stimulated a search for intermediate phases in SBT. Onodera et. al. [7] proposed the existence of an intermediate phase based on the thermal behavior and x-ray diffraction patterns of Bi-rich $Sr_{0.8}Bi_{2.2}Ta_2O_9$ thin films. Structural studies in polycrystalline samples showed the existence of the intermediate phase (occurring between ~ 350 and 550°C), although its space group (*Fmmm* [8], *Amam* [9,10] or *B2cb* [11]) was under debate. A direct observation of ferroelastic domains which disappears about 550 °C supports the ferroelasticity and the *Amam* symmetry of the intermediate phase [12]. Surprisingly, $SrBi_2Nb_2O_9$ shows a single crystallographic phase transition [13] from *A2₁am* to *I*4/*mmm*, rather than the proceeding via the intermediate phase, despite being isomorphous with SBT. Intermediate phases have not been detected either in $SrBi_4Ti_4O_{15}$ and $Bi_4Ti_3O_{12}$ which seem to exhibit a direct phase transition from the high-temperature tetragonal to a ferroelectric phase. However, due to the limitations of the powder method for diffraction analysis, which is the only method used up to now at high temperatures, it cannot be ruled out that intermediate phases indeed exist. Recently, it was shown that the ferroelectric transition in BIT involves the interplay of six different normal modes belonging to four different irreducible representations, indicating that the apparent absence of intermediate phases remains to be explained [2].

Given this experimental background and the difficulties intrinsic to high-temperature measurements, theoretical and simulation studies of the phase transitions sequence in Aurivillius compounds are highly desirable. While first-principles calculations have contributed to the understanding of the electronic structure and structural instabilities of SBT [4,5,14] and other Aurivillius [2,15], these methods are restricted to studying zero-temperature properties. Nowadays, the combination of first-principles calculations with effective Hamiltonian [16], or shell-model [17] techniques offers a multiscale approach to investigate the various functional properties of ferroelectric oxides. Both methods are able to correctly reproduce the phase sequences in $ABO_3$ perovskites although the transition temperatures have relatively large errors (for comparison with experiment, it is customary to rescale the temperature to match the experimental $T_c$). Shell-models directly fitted on first-principles calculations, combined with molecular dynamics simulations revealed powerful to predict the qualitative behavior of pure compounds and alloys [18], including nanostructures [19]. The atomistic simulation of Aurivillius compounds is a theoretical challenge, not only for the complexity of the crystal structure, which is highly anisotropic, but also for the delicate interplay between polar and nonpolar instabilities. In this letter we investigate the thermal sequence of phase transitions in SBT using a shell-model approach with parameters fitted to first-principles calculations.

The model used in the present work contains fourth-order core-shell couplings, long-range Coulombic interactions and short-range interactions described by two different types of potentials. A Rydberg potential $V(r)=(A+Br)\ e^{-r/\rho}$ is used for the Sr-O, Ta-O, and Bi-O pairs, and a Buckingham potential $V(r)=A\ e^{-r/\rho} - C/r^6$ is used for O-O interactions. For more details about the shell model approach see Ref. [17]. The model parameters were fitted completely to first principle results. That is, no explicit experimental data had been used as input. The Local Density Approximation (LDA) calculations were performed with the WIEN2K code [20]. See Ref. [4] for details about the implementation for SBT. The input information for the parameter least-square fit procedure corresponded to LDA results of the optimized crystal structure for the orthorhombic and tetragonal phases, energy as function of volume and strain (c/a, c/b, b/a), underlying potential energy surfaces for structural distortions (ferroelectric and ferrodistortive) from the tetragonal phase, and forces between atoms. The resulting shell model parameters are shown in Table I. We use molecular-dynamics (MD) simulations to determine the phase transition sequence and temperature-dependent properties of the material. The simulations were carried out using the DL-POLY package [21]. The runs were performed employing a Hoover constant-($\sigma$,T) algorithm. Periodic boundary conditions over 7168 atoms were considered.

The resulting model is able to reproduce correctly the T=0K structure of SBT, in good agreement with LDA. It gives an orthorhombic ground state (space group *A21am*) with lattice parameters *a*= 5.46Å, *b*=5.44 Å, *c*= 24.42Å, while the LDA results are a=5.49Å, b=5.48Å and c=24.70 Å. As it is well known, LDA underestimates volume and volume-dependent properties with respect to experimental values. For SBT the experimental lattice constants are *a*=5.52Å, *b*=5.52 Å and *c*=25.02 Å [22]. So, the LDA underestimation of the static structural properties will be translated via the adjusted model to the finite temperature behavior. Regarding the internal atomic positions, the relative coordinates of the orthorhombic *A21am* structure obtained by the model agree with experimental values better than 2%. The spontaneous polarization, however, is underestimated. The model gives a spontaneous polarization $P \approx 8$ μC/cm$^2$ along the a axis. Although this value agree quite well with the polarization measured in SBT ceramics and thin films, it is lower than the value $P = 20$ μC/cm$^2$ measured in single crystals [23]. The absence of ferroelectricity along the [001] is in agreement with experiments.

Three order parameters were defined to characterize the transitions. They are related to the main distortions leading to the low-temperature phase from the tetragonal structure: the spontaneous polarization (P) along the *a* axis and the rotation angles of the TaO$_6$ octahedra around the *a* ($\phi_a$) and *c* ($\phi_c$) axes. The resulting lattice constants and the behavior of the three order parameters as a function of temperature are shown in Figure 1 and 2, respectively. Discontinuities are clearly discernible in the plots of *a, b* and *c* parameters vs. T, suggesting the existence of two phase transitions, one at $T_{c1} \sim 870K$ and another at $T_{c2} \sim 1300$ K. At low temperatures (T < $T_{c1}$) the structure is orthorhombic ($a \neq b \neq c$) and P, $\phi_a$ and $\phi_c$ are all clearly different from zero, indicating the presence of the *A21am* ferroelectric phase. The lattice parameters show strong discontinuities at $T_{c1}$, despite the fact that the structure is still orthorhombic up to $T_{c2}$. In this temperature range ($T_{c1}$ < T < $T_{c2}$) the lattice parameter *a* is slightly different from *b*; P and $\phi_c$ are very close to zero while $\phi_a$ remains finite. This indicates that a transition from the ferroelectric to an orthorhombic paraelectric phase takes place at $T_{c1}$. This transition involves not only loss of polarization, but also the loss of the $TaO_6$ octahedral tilt mode around the *c* axis. The orthorhombic paraelectric structure is then characterized by the rotation of the $TaO_6$ octahedra around the a-axis (*Amam* symmetry). At $T_{c2}$, the lattice constant *c* shows a marked change of slope, *a* and *b* take the same value, and $\phi_a$ vanishes. This is an indication that a transition to a tetragonal paralectric *I4/mmm* phase occurs at this temperature. Regarding the order of the transitions, in the inset of Figure 2 we plot the unit cell volume (V = *a.b.c*) as a function of temperature. The strong volume anomaly observed at the transition temperature $T_{c1}$, together with the discontinuous change of the order parameter P, suggest that the ferroelectric phase transition is first order. On the contrary, no volume anomaly is observed at $T_{c2}$. This fact and the continuous change of the order parameter $\phi_a$ across the transition indicate that the transition from the intermediate to the tetragonal phase is second order. This is in agreement with the results of Kamba et.al. who observed a gradual disappearance of elastic domains on heating, suggesting that the transition at $T_{c2}$ is near second order [12].

One advantage of our approach is the possibility to calculate single crystal properties. The experimental studies of Aurivillius compounds are carried out mainly on ceramics and thin films and rarely on single crystals due to the lack of crystals of sufficiently good quality. Single crystal results are important to determine the anisotropy of material properties. For instance, we used the atomistic model to investigate the dielectric properties of SBT through the calculation of the dielectric constant ($\epsilon$) along crystallographic directions. The temperature dependence of $\epsilon$ is determined by calculating the change in the polarization under an applied electric field. Figure 3 shows the dielectric constants $\epsilon_{11}$, $\epsilon_{22}$, and $\epsilon_{33}$ as function of temperature. The *a, b* and *c* axes of the orthorhombic phase lie along the 1, 2 and 3 directions, respectively. The behavior showed in Figure 3 is in qualitative agreement with experimental measurements in single crystals [23], although in the latter the in-plane anisotropy ($\epsilon_{11}$ and $\epsilon_{22}$) was not measured. The maximum of $\epsilon_{11}$ corresponding to the ferro-paralectric phase transition is observed at $T_{c1}$. No anomalies are observed in $\epsilon_{22}$ and $\epsilon_{33}$ at this temperature, which confirms that the small peak observed experimentally for $\epsilon_{33}$ around $T_{c1}$ [23] is of extrinsic character. Although no dielectric anomalies are observed at $T_{c2}$, the dielectric behavior provides an alternative way for the identification of the intermediate phase. While $\epsilon_{11} = \epsilon_{22}$ in the tetragonal phase (T>$T_{c2}$), $\epsilon_{11}$ is considerable larger than $\epsilon_{22}$ in the intermediate phase ($T_{c1}$ < T < $T_{c2}$). That difference arises from the atomic displacements along the *b* direction produced by the rotation of $TaO_6$ octahedra around the a-axis. The anisotropy of the dielectric response in the *a-b* plane can be used for the experimental detection of intermediate phases in other Aurivillius compounds, provided the existence of single crystals of sufficient quality.

In summary, we have shown that the combination of first-principles calculations with shell-model techniques offers a multiscale approach to investigate finite temperature properties of Aurivillius compounds. In particular, we demonstrated the existence of an intermediate paraelectric phase in SBT without using any explicit experimental data as input; this phase naturally emerges from the simulated phase diagram. This approach is a powerful tool to investigate the presence of intermediate phases, not yet observed due to experimental difficulties, in other Aurivillius compounds.

We thank R. Migoni for useful discussions. This work was sponsored by Consejo Nacional de Investigaciones Científicas y Tecnológicas de la Republica Argentina (CONICET). MGS thanks support from CIUNR.

Table I: Shell model parameters for SBT. Units of energy, length, and charge are given in eV, Å and electrons respectively.

| Atom | Core charge | Shell charge | $k_2$ | $k_4$ |
|---|---|---|---|---|
| Sr | -2.060 | 3.346 | 45.0 | 6464.8 |
| Bi | 8.575 | -5.760 | 269.7 | 63.3 |
| Ta | 10.380 | -5.492 | 1064.8 | 32.9 |
| O | 0.419 | -2.274 | 17.2 | 1079.3 |
| Short-range | A | B | $\rho$ | C |
| Sr-O | 872.043 | -67.565 | 0.362147 | 0.0 |
| Bi-O | 24135.937 | -9301.572 | 0.291147 | 0.0 |
| Ta-O | 4681.849 | -689.125 | 0.290416 | 0.0 |
| O-O | 17585.281 | 0.0 | 0.239138 | 23.226 |

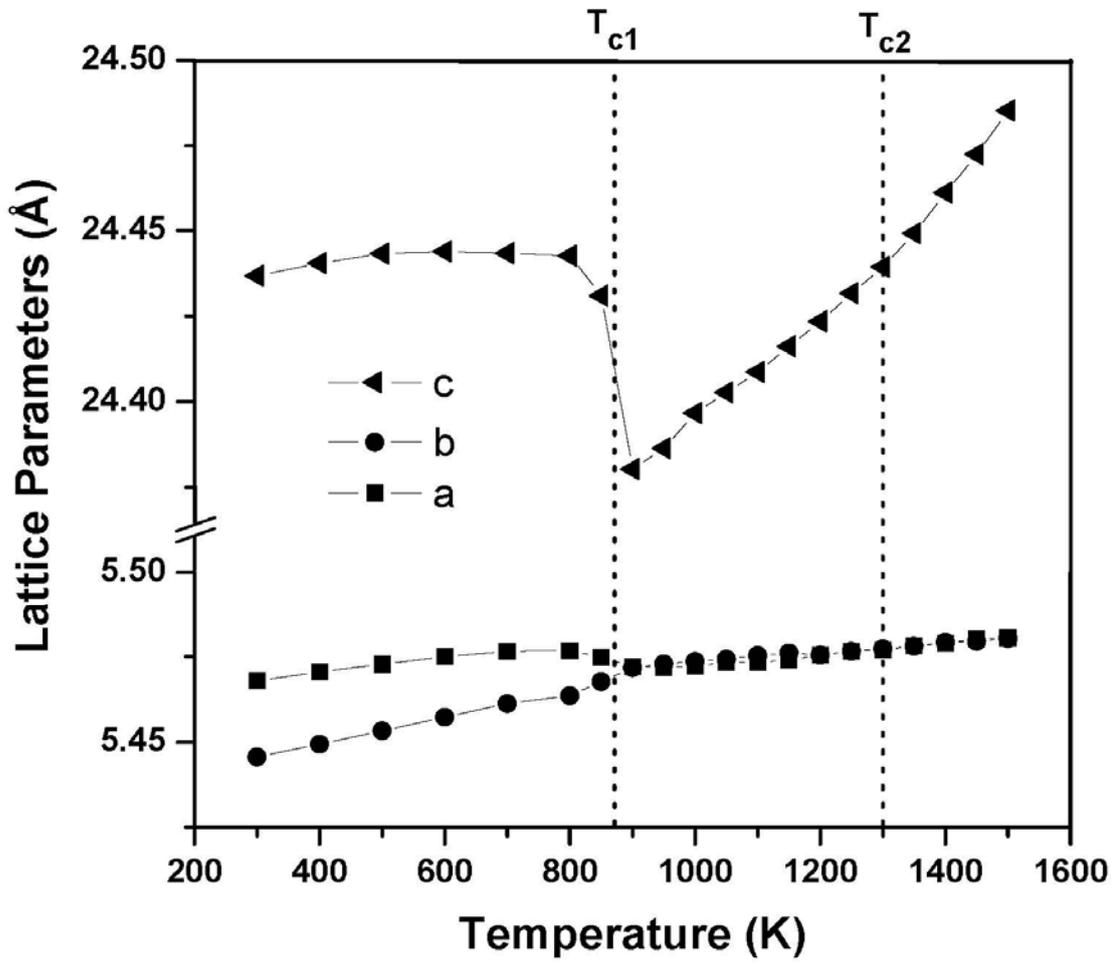

Figure 1: Thermal evolution of lattice parameters for $SrBi_2Ta_2O_9$ determined from MD simulations. The vertical lines indicate the approximate positions of the structural phase transitions.

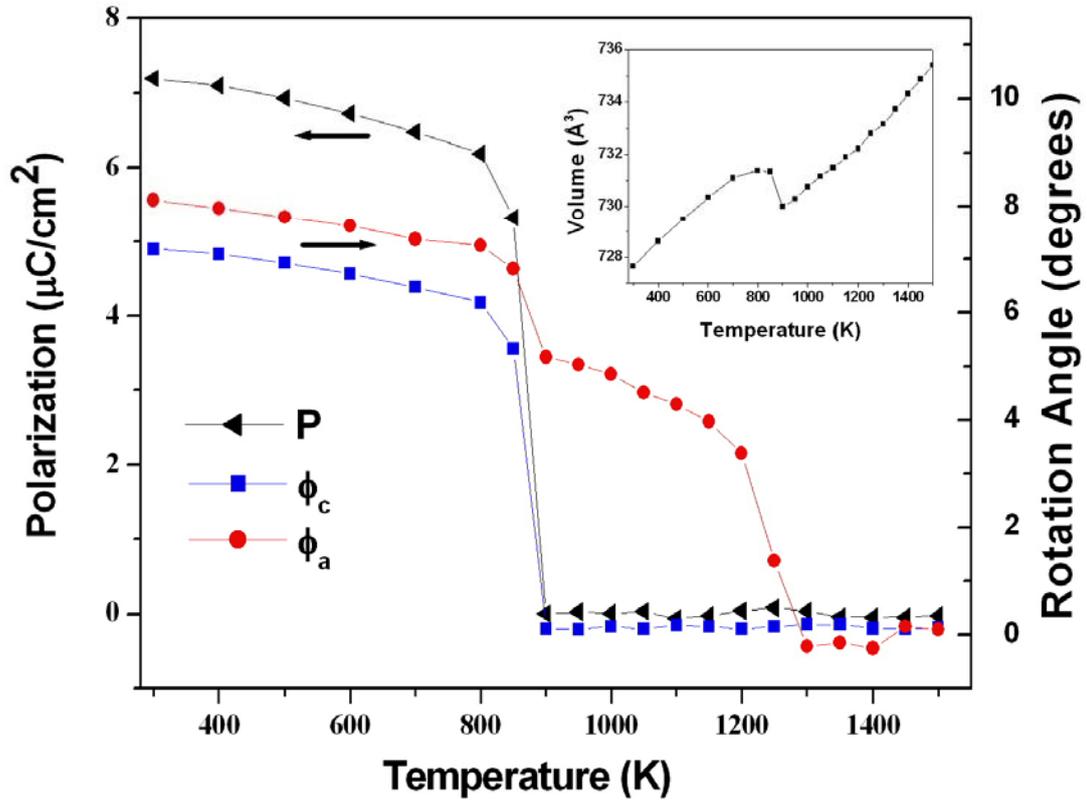

Figure 2 (color on line): Thermal evolution of the order parameters for $SrBi_2Ta_2O_9$ determined from MD simulations. P is the spontaneous polarization along the *a* axis. $\phi_a$ and $\phi_c$ are the rotation angles of the $TaO_6$ octahedra around the *a* and *c* axes, respectively. The inset shows the temperature dependence of the conventional unit cell volume.

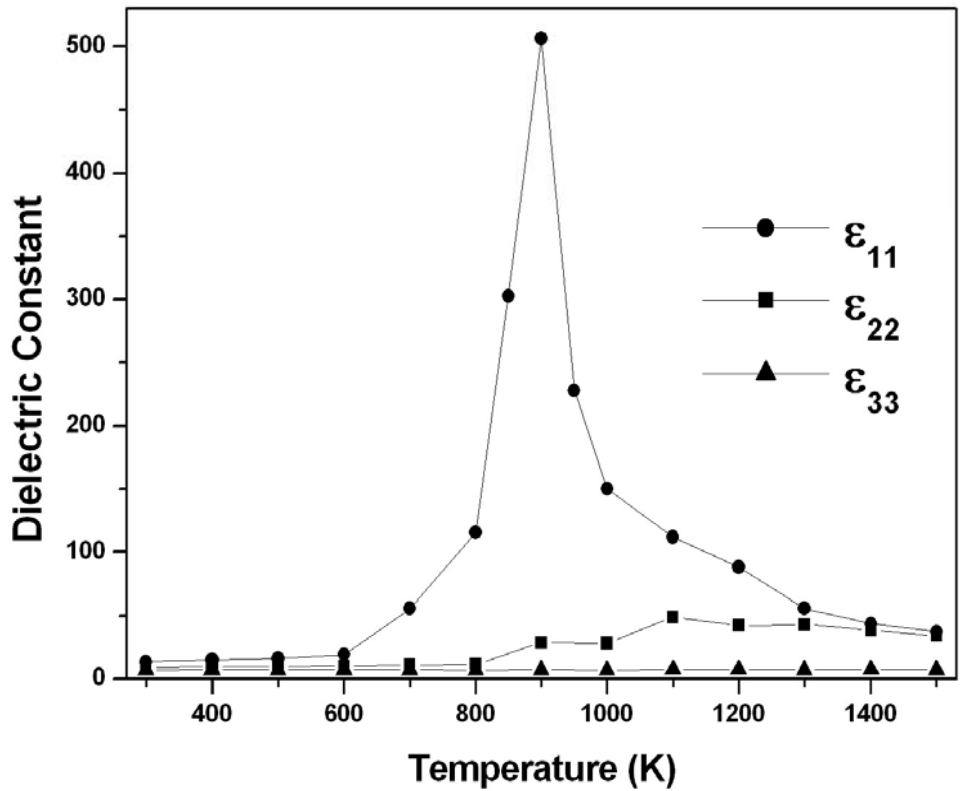

Figure 3: Temperature dependence of the dielectric permittivity along the crystallographic directions for $SrBi_2Ta_2O_9$ determined from MD simulations. The *a*, *b* and *c* axes of the orthorhombic phase lie along the 1, 2 and 3 directions, respectively.